\documentclass[a4paper,11pt]{article}

\usepackage{jheppub_mod} 
\usepackage[T1]{fontenc} 

\usepackage{amsmath,booktabs}
\usepackage{graphicx}
\usepackage{xspace}
\usepackage{color}
\pagecolor{white}
\usepackage[dvipsnames]{xcolor}
\usepackage{url}
\usepackage[utf8]{inputenc}
\usepackage{epstopdf}
\usepackage{hyperref}
\usepackage{multirow}

\newcommand{\mpi}{M_\pi}
\newcommand{\mpii}{M_{\pi^0}}
\newcommand{\beq}{\begin{equation}}
\newcommand{\eeq}{\end{equation}}

\newcommand{\eps}{\epsilon}

\newcommand{\Order}{\mathcal{O}}

\newcommand{\mw}{M_\omega}
\newcommand{\Gw}{\Gamma_\omega}

\newcommand{\epsrw}{\eps_{\omega}}
\newcommand{\amurw}{a_\mu^{\rho\text{--}\omega}}

\newcommand{\mr}{M_\rho}

\newcommand{\GeV}{\,\text{GeV}}
\newcommand{\MeV}{\,\text{MeV}}

\renewcommand{\Im}{\text{Im}\,}
\renewcommand{\Re}{\text{Re}\,}
\newcommand{\disc}{\text{disc}\,}

\allowdisplaybreaks[1]

\preprint{PSI-PR-22-26,  ZU-TH 41/22}

\title{Isospin-breaking effects in the two-pion contribution to hadronic vacuum polarization}

\author[a]{Gilberto Colangelo,}
\author[a]{Martin Hoferichter,}
\author[b]{Bastian Kubis,}
\author[c,d]{and Peter Stoffer}

\affiliation[a]{Albert Einstein Center for Fundamental Physics, Institute for Theoretical Physics, University of Bern, Sidlerstrasse 5, 3012 Bern, Switzerland}

\affiliation[b]{Helmholtz-Institut f\"ur Strahlen- und Kernphysik (Theorie) and \\
Bethe Center for Theoretical Physics, Universit\"at Bonn, 53115 Bonn, Germany}

\affiliation[c]{Physik-Institut, Universit\"at Z\"urich, Winterthurerstrasse 190, 8057 Z\"urich, Switzerland}

\affiliation[d]{Paul Scherrer Institut, 5232 Villigen PSI, Switzerland}

\emailAdd{gilberto@itp.unibe.ch}
\emailAdd{hoferichter@itp.unibe.ch}
\emailAdd{kubis@hiskp.uni-bonn.de}
\emailAdd{stoffer@physik.uzh.ch}

\abstract{Isospin-breaking (IB) effects in the two-pion contribution to hadronic vacuum polarization (HVP) can be resonantly enhanced, if related to the interference of the $\rho(770)$ and $\omega(782)$ resonances. This particular IB contribution to the pion vector form factor and thus the line shape in $e^+e^-\to \pi^+\pi^-$ can be described by the residue at the $\omega$ pole---the $\rho$--$\omega$ mixing parameter $\epsrw$. Here, we argue that while in general analyticity requires this parameter to be real, the radiative channels $\pi^0\gamma$, $\pi\pi\gamma$, $\eta\gamma$ can induce a small phase, whose size we estimate as $\delta_\eps=3.5(1.0)^\circ$ by using a narrow-width approximation for the intermediate-state vector mesons. We then perform fits to the $e^+e^-\to \pi^+\pi^-$ data base and study the consequences for the two-pion HVP contribution to the anomalous magnetic moment of the muon, its IB part due to $\rho$--$\omega$ mixing, and the mass of the $\omega$ resonance. We find that the global fit does prefer a non-vanishing value of $\delta_\eps=4.5(1.2)^\circ$, close to the narrow-resonance expectation, but with a large spread among the data sets,
indicating systematic differences in the $\rho$--$\omega$ region.}

\begin{document}
\maketitle
	
\section{Introduction}
\label{sec:intro}

The two-pion channel gives the dominant contribution to hadronic vacuum polarization (HVP) in the low-energy region most relevant for the anomalous magnetic moment of the muon~\cite{Aoyama:2020ynm}, adding about $70\%$ of the total leading-order (LO) effect~\cite{Davier:2017zfy,Keshavarzi:2018mgv,Colangelo:2018mtw,Hoferichter:2019gzf,Davier:2019can,Keshavarzi:2019abf}
\beq
a_\mu^\text{HVP, LO}\big|_{e^+e^-}=693.1(4.0)\times 10^{-10}.
\label{HVPee}
\eeq
Its contribution needs to be understood at a level of at least $0.3\%$ to match the final precision expected from the Fermilab E989 experiment~\cite{Muong-2:2015xgu}. The nominal combined sensitivity of the $2\pi$ data sets entering Eq.~\eqref{HVPee}---from SND~\cite{Achasov:2005rg,Achasov:2006vp}, CMD-2~\cite{Akhmetshin:2001ig,Akhmetshin:2003zn,Akhmetshin:2006wh,Akhmetshin:2006bx}, BESIII~\cite{Ablikim:2015orh}, CLEO~\cite{Xiao:2017dqv}, and dominated by the  precision data sets from 
BaBar~\cite{Aubert:2009ad,Lees:2012cj} and KLOE~\cite{Ambrosino:2008aa,Ambrosino:2010bv,Babusci:2012rp,Anastasi:2017eio}---does reach $0.4\%$, but becomes diluted due to a tension between BaBar and KLOE, inflating the $2\pi$ uncertainty included in Eq.~\eqref{HVPee} to $0.7\%$. In dispersive approaches~\cite{Colangelo:2018mtw,Ananthanarayan:2018nyx,Davier:2019can,Hoferichter:2019gzf,Stamen:2022uqh} also space-like data~\cite{Dally:1982zk,Amendolia:1986wj} can be used, and while stabilizing the extrapolation to the space-like region, their impact on the time-like HVP integral is minor. More recently, new data from
SND~\cite{SND:2020nwa} have become available, lying in between BaBar and KLOE, but not at a comparable level of precision that would allow one to resolve the tension. Such new precision measurements are expected  
from CMD-3~\cite{Ryzhenenkov:2020vrk}, BaBar~\cite{Abbiendi:2022liz}, BESIII~\cite{BESIII:2020nme}, and Belle~II~\cite{Belle-II:2018jsg} in the future. 

Improved understanding of the $2\pi$ channel has further become critical to address the emerging tension between lattice QCD~\cite{Borsanyi:2020mff,Ce:2022kxy,Alexandrou:2022amy,Davies:2022epg} and $e^+e^-$ data at least for the intermediate window quantity~\cite{Blum:2018mom}, 
with immediate consequences 
for the current $4.2\sigma$
discrepancy for the anomalous magnetic moment of the muon between experiment~\cite{Muong-2:2006rrc,Muong-2:2021ojo,Muong-2:2021ovs,Muong-2:2021xzz,Muong-2:2021vma} and the prediction in the Standard Model~\cite{Aoyama:2020ynm,Aoyama:2012wk,Aoyama:2019ryr,Czarnecki:2002nt,Gnendiger:2013pva,Davier:2017zfy,Keshavarzi:2018mgv,Colangelo:2018mtw,Hoferichter:2019gzf,Davier:2019can,Keshavarzi:2019abf,Hoid:2020xjs,Kurz:2014wya,Melnikov:2003xd,Masjuan:2017tvw,Colangelo:2017qdm,Colangelo:2017fiz,Hoferichter:2018dmo,Hoferichter:2018kwz,Gerardin:2019vio,Bijnens:2019ghy,Colangelo:2019lpu,Colangelo:2019uex,Blum:2019ugy,Colangelo:2014qya} when the HVP contribution is derived from $e^+e^-\to\text{hadrons}$ cross-section data. While the detailed comparison to lattice QCD as well as related observables defines an important path forward~\cite{Passera:2008jk,Crivellin:2020zul,Keshavarzi:2020bfy,Malaescu:2020zuc,Colangelo:2020lcg,Ce:2022eix,Colangelo:2022jxc,Colangelo:2022vok}, so does renewed scrutiny of the data-driven approach.  

For the $2\pi$ channel, new precision data sets constitute the clear first priority, but another aspect concerns the role of radiative corrections~\cite{Abbiendi:2022liz,WorkingGrouponRadiativeCorrections:2010bjp}, in particular, the question in which cases the  use of a point-like approximation~\cite{Hoefer:2001mx,Czyz:2004rj,Gluza:2002ui,Bystritskiy:2005ib,Campanario:2019mjh} for the pion might miss relevant effects, as recently observed in the forward--backward asymmetry~\cite{Ignatov:2022iou,Colangelo:2022lzg}, and currently under study for the $C$-even contributions~\cite{JMPhDThesis}. In this work, we study a complementary point, i.e., not isospin-breaking (IB) effects that manifest themselves as final- or initial-state radiation, but corrections that are typically absorbed into the pion vector form factor (VFF) itself.\footnote{Such corrections were studied before in the context of relating VFF measurements in $e^+e^-\to\pi^+\pi^-$ to $\tau^\pm\to\pi^\pm\pi^0\nu_\tau$ data~\cite{Alemany:1997tn,Cirigliano:2001er,Cirigliano:2002pv,Davier:2010fmf,Jegerlehner:2011ti}. Here, we aim instead at a rigorous implementation of $\rho$--$\omega$ mixing in a dispersive framework, both to quantify its impact on $a_\mu^\text{HVP, LO}$ and as another powerful consistency check on the $e^+e^-$ data base.} The most prominent such correction arises from $\rho$--$\omega$ mixing. From a dispersive point of view the fact that the $\omega(782)$ resonance is so narrow allows one to  describe this interference in terms of a single real parameter: the $\rho$--$\omega$ mixing parameter $\epsrw$.
But given the extraordinary precision requirements for the $2\pi$ channel together with the resonance enhancement in the $\rho$--$\omega$ region, even higher-order effects may affect the value of this parameter and generate non-negligible effects in $a_\mu^\text{HVP, LO}$.  Most notably, the radiative channels $\pi^0\gamma$, $\pi\pi\gamma$, $\eta\gamma$, all of which couple to both $\rho$ and $\omega$, can induce imaginary parts in the mixing and thereby an effective small phase $\delta_\eps$ in the parameter $\epsrw$ to which $e^+e^-$ data might be sensitive.

To derive the phenomenological consequences of this phase we first generalize the dispersive representation of the pion VFF from Ref.~\cite{Colangelo:2018mtw} and estimate its size based on a narrow-resonance approach, see Sec.~\ref{sec:phase}. We then perform fits, to individual data sets and globally, allowing for a free phase $\delta_\eps$, to assess consistency both among the data sets and with the narrow-resonance expectation, see Sec.~\ref{sec:fits}. Consequences for the IB contribution to $a_\mu$ due to $\rho$--$\omega$ mixing and the $\omega$ mass are discussed in Secs.~\ref{sec:IB} and~\ref{sec:Momega}, respectively, before concluding in Sec.~\ref{sec:conclusions}.

\section{Radiative channels and phase in the $\boldsymbol{\rho}$--$\boldsymbol{\omega}$ mixing parameter}
\label{sec:phase}

\subsection{Dispersive representation}

Dispersive representations for the pion VFF, $F_\pi^V(s)$, that link the matrix element of the electromagnetic current $j_\text{em}^\mu=(2\bar u\gamma^\mu u-\bar d\gamma^\mu d-\bar s \gamma^\mu s)/3$,
 \beq
	\langle \pi^\pm(p') | j_\mathrm{em}^\mu(0) | \pi^\pm(p) \rangle =\pm (p'+p)^\mu F_\pi^V((p'-p)^2),
 \eeq
to $\pi\pi$ scattering
have been used for a long time in the literature~\cite{DeTroconiz:2001rip,Leutwyler:2002hm,Colangelo:2003yw,deTroconiz:2004yzs,Hanhart:2012wi,Ananthanarayan:2013zua,Ananthanarayan:2016mns,Hoferichter:2016duk,Hanhart:2016pcd,Colangelo:2018mtw,Ananthanarayan:2018nyx,Davier:2019can,Colangelo:2021moe}, 
not only for the HVP application, but also for hadronic light-by-light scattering (HLbL), where the extrapolation into the space-like region enters~\cite{Hoferichter:2013ama,Colangelo:2014dfa,Colangelo:2014pva,Colangelo:2015ama,Colangelo:2017qdm,Colangelo:2017fiz}.

Here, we build upon the representation from Ref.~\cite{Colangelo:2018mtw} (in turn based on Refs.~\cite{Leutwyler:2002hm,Colangelo:2003yw}), whose main features can be summarized as follows: the VFF is decomposed into three factors   
\beq
	\label{eq:PionVFF}
	F_\pi^V(s) = \Omega_1^1(s) G_\omega(s) G_\text{in}^N(s),
\eeq
corresponding to $2\pi$, $3\pi$, and higher intermediate states, respectively. The Omn\`es function~\cite{Omnes:1958hv}
\begin{align}
	\label{eq:OmnesFunction}
	\Omega_1^1(s) = \exp\left\{ \frac{s}{\pi} \int_{4\mpi^2}^\infty ds^\prime \frac{\delta_1^1(s^\prime)}{s^\prime(s^\prime-s)} \right\}
\end{align}
implements $2\pi$ singularities in terms of the 
isospin $I=1$ elastic $\pi\pi$ phase shift $\delta_1^1(s)$ in the isospin limit. The phase shift is further constrained by $\pi\pi$ Roy equations~\cite{Roy:1971tc}, which are solved with the phase shifts at $s_0=(0.8\GeV)^2$ and $s_1=(1.15\GeV)^2$ as free parameters~\cite{Ananthanarayan:2000ht,Caprini:2011ky}. Systematic errors from the asymptotic continuation of $\delta^1_1$ beyond $s_1$ are treated as described in Ref.~\cite{Colangelo:2018mtw}.

The focus of this work is the second factor, $G_\omega$, which takes into account the effect of $3\pi$ intermediate states. Here, the parameterization from Ref.~\cite{Colangelo:2018mtw} reads
\beq
\label{Gomega}
	G_\omega(s) = 1 + \frac{s}{\pi} \int_{9\mpi^2}^\infty ds^\prime \frac{\Im g_\omega(s^\prime)}{s^\prime(s^\prime-s)} \left( \frac{1 - \frac{9\mpi^2}{s^\prime}}{1 - \frac{9\mpi^2}{\mw^2}} \right)^4,
\eeq
with
\beq
\label{gomega}
	g_\omega(s) = 1 + \epsilon_\omega \frac{s}{(\mw - \frac{i}{2} \Gw)^2 - s}.
\eeq
The dispersive reconstruction in Eq.~\eqref{Gomega} ensures both the absence of unphysical imaginary parts below $s=9\mpi^2$ (for real $\epsrw$) and the correct threshold behavior above~\cite{Leutwyler:2002hm}. In this formulation, $\epsrw$ is an effective parameter tightly related to the residue at the $\omega$ pole. The latter, however, is complex in general, but its phase is expected to be tiny: $\delta_\eps\simeq \arctan \Gw/\mw\simeq 0.6^\circ$ arising from the analytic continuation from the real axis to the pole position in the complex plane. Such a small difference is of no concern and if we take it as a measure of the systematic uncertainty in the phase of $\epsrw$, allows us to view the latter as the residue at the $\omega$ pole. With the threshold behavior and the pole parameters determined, the resulting $G_\omega(s)$ is then largely insensitive to the parameterization of $g_\omega(s)$, e.g., the numerator could be taken to a constant without any relevant changes to the fit outcome. The main observation in this paper is that the assumption of a real $\epsrw$ no longer holds if further IB effects due to radiative channels, $X=\pi^0\gamma, \pi\pi\gamma, \eta\gamma, \ldots$, are considered, and these imaginary parts, despite being small, can alter the fit parameters in a significant way, as only the modulus $|F_\pi^V(s)|^2$ is probed by the fit to the cross-section data. 

Finally, for the inelastic channels we continue to use a conformal polynomial, whose phase is constrained by the Eidelman--\L{}ukaszuk bound~\cite{Lukaszuk:1973jd,Eidelman:2003uh}. Its threshold is chosen as the $\omega\pi^0$ threshold, below which inelasticities are negligibly small. After removing the $S$-wave cusp, $G_\text{in}^N(s)$ involves $N-1$ free parameters, which together with $\epsrw$ and $\delta_1^1(s_0)$, $\delta_1^1(s_1)$ are to be constrained in the fit. The fit range is restricted to $s\leq 1 \GeV^2$, going beyond would require including the effects of $\rho'$, $\rho''$ resonances along the lines of Refs.~\cite{Hanhart:2012wi,Chanturia:2022rcz}, an extension left for future work.

\subsection{Radiative channels}
\label{sec:radiative}

Both $\rho$ and $\omega$ possess non-negligible branching fractions into radiative channels, of which $\pi^0\gamma$ yields the largest contribution. The corresponding imaginary part in the pion VFF can be expressed as
\beq
\label{FpiV_f1F}
\Im F_\pi^V(s)\big|_{\pi^0\gamma}=\frac{\alpha(s-\mpii^2)^3}{48s}F_{\pi^0\gamma^*\gamma^*}(s,0)\big(f_1(s)\big)^*,
\eeq
where $F_{\pi^0\gamma^*\gamma^*}$ is the pion transition form factor normalized according to
\beq
F_{\pi^0\gamma^*\gamma^*}(0,0)=F_{\pi\gamma\gamma}=\sqrt{\frac{4\Gamma[\pi^0\to\gamma\gamma]}{\pi\alpha^2\mpii^3}}
\eeq
and $f_1$ denotes the $P$-wave projection of the $\gamma\pi\to\pi\pi$ amplitude, see Refs.~\cite{Hoferichter:2012pm,Schneider:2012ez,Hoferichter:2014vra,Hoferichter:2017ftn,Niehus:2021iin} for detailed discussions of these amplitudes.  
To map this imaginary part onto $\Im \epsrw$, we first write the full pion VFF in the approximation
\beq
\label{FpiV_NW}
F_\pi^V(s)= \bigg(1+\epsrw\frac{s}{\mw^2-s-i\eps}\bigg)\Omega^1_1(s),
\eeq
where we have neglected inelastic corrections for the time being to focus on the interplay of $\rho$ and $\omega$ resonances, with the $\omega$ approximated in the narrow-width limit for simplicity. As a first step, we show that an imaginary part in $\epsrw$ in this form is actually compatible with unitarity. Applying Cutkosky rules to Eq.~\eqref{FpiV_NW}, we have
\begin{align}
\label{FpiV_Im}
 \frac{1}{2i}\disc F_\pi^V(s)\big|_{2\pi}&= \bigg(1+\epsrw\frac{s}{\mw^2-s-i\eps}\bigg) \Im \Omega^1_1(s),\notag\\
  \frac{1}{2i}\disc F_\pi^V(s)\big|_{3\pi}&=\epsrw^* s\pi\delta(s-\mw^2)\big(\Omega^1_1(s)\big)^*,\notag\\
  \frac{1}{2i}\disc F_\pi^V(s)\big|_{\pi^0\gamma}&= \Im \epsrw \frac{s}{\mw^2-s-i\eps}\big(\Omega^1_1(s)\big)^*.
\end{align}
These equations are only consistent as long as the sum of these discontinuities is purely imaginary. Collecting all terms, this consistency check is indeed satisfied,
\begin{align}
 &\Im\bigg[\frac{1}{2i}\disc F_\pi^V(s)\big|_{2\pi}+\frac{1}{2i}\disc F_\pi^V(s)\big|_{3\pi}+\frac{1}{2i}\disc F_\pi^V(s)\big|_{\pi^0\gamma}\bigg]\notag\\
 &=\Im\Omega^1_1(s)\bigg(\Im\epsrw \frac{s}{\mw^2-s}+\Re \epsrw s\pi\delta(s-\mw^2)\bigg)\notag\\
 &+s \pi\delta(s-\mw^2)\bigg(-\Re\epsrw\Im\Omega^1_1(s)-\Re\Omega^1_1(s)\Im\epsrw\bigg)\notag\\
 &+ \Im \epsrw \bigg(-\frac{s}{\mw^2-s}\Im \Omega^1_1(s)+\Re\Omega^1_1(s)s\pi\delta(s-\mw^2)\bigg)\notag\\
 &=0,
\end{align}
so that as long as imaginary parts are avoided below the respective thresholds a phase in $\epsrw$ is indeed possible. 

Next, the comparison of Eqs.~\eqref{FpiV_f1F} and~\eqref{FpiV_Im} points towards a strategy for a practical implementation, with $\Omega_1^1(s)$ corresponding to $f_1(s)$, and the $\omega$ propagator to the $\omega$ contribution in $F_{\pi^0\gamma^*\gamma^*}(s,0)$. 
  The latter is given by 
  \beq
  F_{\pi^0\gamma^*\gamma^*}(s,0) \simeq \frac{g_{\omega\pi\gamma}}{g_{\omega\gamma}}\frac{\mw^2}{\mw^2-s-i\eps}
  \eeq
  (see, e.g., Ref.~\cite{Zanke:2021wiq}), while the former can be approximated by~\cite{Hoferichter:2017ftn}
  \beq
  f_1(s) \simeq \frac{2 g_{\rho\pi\gamma} g_{\rho\pi\pi}}{\mr^2-s-i\mr\Gamma_\rho},\qquad
  \Omega_1^1(s)\simeq \frac{\mr^2}{\mr^2-s-i\mr\Gamma_\rho},
  \eeq
  leading to
  \beq
  \Im\epsrw\simeq \frac{\alpha(s-\mpii^2)^3}{24s} \frac{g_{\omega\pi\gamma} g_{\rho\pi\gamma} g_{\rho\pi\pi}}{g_{\omega\gamma} \mr^2} . \label{Imepsrw_couplings}
  \eeq
  Inserting the expressions for the radiative decay widths,
  \beq
  \Gamma[V \to \pi^0\gamma] = \frac{\alpha(M_V^2-\mpii^2)^3}{24M_V^3}|g_{V\pi\gamma}|^2, \qquad V=\omega,\rho,
  \eeq
  as well as the VMD predictions  $g_{\rho\pi\pi}=g_{\rho\gamma}=g_{\omega\gamma}/3$, and evaluating Eq.~\eqref{Imepsrw_couplings} 
at $s=M_V^2\simeq M_\rho^2\simeq \mw^2$, we find
\beq
\Im\epsrw \simeq \frac{\sqrt{\Gamma[\omega\to \pi^0\gamma]\Gamma[\rho\to \pi^0\gamma]}}{3M_V}.
\eeq
In fact, in the narrow-width limit the same relation can be established for an arbitrary intermediate state $X$, leading to estimates for the phases 
 around $2.8^\circ$ ($\pi^0\gamma$), $0.2^\circ$ ($\eta\gamma$), and $0.02^\circ$ ($\pi^0\pi^0\gamma$) when using the averages from Ref.~\cite{ParticleDataGroup:2020ssz} for branching fractions and masses. For the charged channel $\pi^+\pi^-\gamma$ one needs to take into account the fact that the presence of the Born-term contribution leads to an infrared divergence, in such a way that branching fractions are typically quoted with a cut $E_\gamma=50\MeV$ in the photon energy~\cite{Dolinsky:1991vq,Moussallam:2013una}. However, combined with virtual corrections calculated in a scalar-QED approximation one can define an infrared-safe decay width as
 \beq
 \Gamma[V\to\pi^+\pi^-\gamma]=\Gamma[V\to\pi^+\pi^-]\frac{\alpha}{\pi}\eta(M_V^2),
 \eeq
 where explicit expressions for the function $\eta$ can be found in Refs.~\cite{Hoefer:2001mx,Czyz:2004rj,Gluza:2002ui,Bystritskiy:2005ib}. This procedure gives an estimate of $0.4^\circ$ for the $\pi^+\pi^-\gamma$ channel, subject to minor corrections from non-Born contributions~\cite{Moussallam:2013una}. As for the relative signs, VMD arguments show that the $\pi^0\gamma$ and $\eta\gamma$ channels enter with the same sign (in the standard phase conventions, both intermediate states couple with the same sign to $\rho$ and $\omega$), and Born-term dominance for $\rho,\omega\to\pi^+\pi^-\gamma$, as well as the positive sign of $\Re\epsrw$, suggest that its contribution should also add to the other two.
 Taking further potential corrections due to the analytic continuation to the $\omega$ pole as the uncertainty, 
 we conclude that the range $\delta_\eps=3.5(1.0)^\circ$ should give a realistic estimate of the phase in $\epsrw$ that can be expected.

 Finally, the discussion of the $\pi^0\gamma$ channel suggests that the dominant threshold can be reproduced by implementing the imaginary part in $\epsrw$ via
\begin{align}
\label{eq:GomegaMod}
	G_\omega(s) = 1 &+ \frac{s}{\pi} \int_{9\mpi^2}^\infty ds^\prime \frac{\Re\epsrw}{s^\prime(s^\prime-s)} \Im\left[ \frac{s'}{(\mw - \frac{i}{2} \Gw)^2 - s'} \right]  \left( \frac{1 - \frac{9\mpi^2}{s^\prime}}{1 - \frac{9\mpi^2}{\mw^2}} \right)^4 \nonumber\\
		&+ \frac{s}{\pi} \int_{\mpii^2}^\infty ds^\prime \frac{\Im\epsrw}{s^\prime(s^\prime-s)} \Re\left[ \frac{s'}{(\mw - \frac{i}{2} \Gw)^2 - s'} \right]  \left( \frac{1 - \frac{\mpii^2}{s^\prime}}{1 - \frac{\mpii^2}{\mw^2}} \right)^3 ,
\end{align}
but we also checked that the fit results are largely insensitive to the details of the implementation, such as the inclusion of the explicit $\pi^0\gamma$ threshold in the unphysical region of the pion VFF. In fact, even replacing $G_\omega(s)$ with $g_\omega(s)$ only leads to small changes as long as the imaginary part in $\epsrw$ is kept, in line with the expectation that it is solely the resonance enhancement that makes these higher-order effects relevant. 

\section{Fits to $\boldsymbol{e^+e^-\to 2\pi}$ data}
\label{sec:fits}

\begin{table}[t]
\small
\centering
\renewcommand{\arraystretch}{0.95}
\scalebox{0.97}{
\begin{tabular}{l c l l l l l l l l}
\toprule
	 & $\chi^2/$dof & $p$-value & $\mw$ [MeV] & $10^3\times \Re\eps_{\omega}$ & $\delta_\eps$ [${}^\circ$] & $10^{10} \times a_\mu^{\pi\pi}|_{\leq 1\GeV}$\\
	\midrule
	SND06 & $1.40$ & $5.3\%$ & $781.49(32)(2)$ & $2.03(5)(2)$ & & $499.7(6.9)(4.1)$\\
	 & $1.08$ & $35\%$ & $782.11(32)(2)$ & $1.98(4)(2)$ & $8.5(2.3)(0.3)$ & $497.8(6.1)(4.9)$\\[0.1cm]
	CMD-2 & $1.18$ & $14\%$ & $781.98(29)(1)$ & $1.88(6)(2)$ & & $496.9(4.0)(2.3)$\\
	 & $1.01$ & $45\%$ & $782.64(33)(4)$ & $1.85(6)(4)$ & $11.4(3.1)(1.0)$ & $495.8(3.7)(4.2)$\\[0.1cm]
	BaBar & $1.14$ & $5.7\%$ & $781.86(14)(1)$ & $2.04(3)(2)$ & & $501.9(3.3)(2.0)$\\
	 & $1.14$ & $5.5\%$ & $781.93(18)(4)$ & $2.03(4)(1)$ & $1.3(1.9)(0.7)$ & $501.9(3.3)(1.8)$\\[0.1cm]
	KLOE & $1.36$ & $7.4 \times 10^{-4}$ & $781.82(17)(4)$ & $1.97(4)(2)$ & & $492.0(2.2)(1.8)$\\
	 & $1.27$ & $6.7 \times 10^{-3}$ & $782.50(25)(6)$ & $1.94(5)(2)$ & $6.8(1.8)(0.5)$ & $491.0(2.2)(2.0)$\\[0.1cm]
	KLOE$''$ & $1.20$ & $3.1\%$ & $781.81(16)(3)$ & $1.98(4)(1)$ & & $491.8(2.1)(1.8)$\\
	 & $1.13$ & $10\%$ & $782.42(23)(5)$ & $1.95(4)(2)$ & $6.1(1.7)(0.6)$ & $490.8(2.0)(1.7)$\\[0.1cm]
	BESIII & $1.12$ & $25\%$ & $782.18(51)(7)$ & $2.01(19)(9)$ & & $490.8(4.8)(3.9)$ \\
	 & \color{gray} $1.02$ & \color{gray} $44\%$ & \color{gray} $783.05(60)(2)$ & \color{gray} $1.99(19)(7)$ & \color{gray} $17.6(6.9)(1.2)$ & \color{gray} $490.3(4.5)(3.1)$ \\[0.1cm]
	SND20 & $2.93$ & $3.3 \times 10^{-7}$ & $781.79(30)(6)$ & $2.04(6)(3)$ & & $494.2(6.7)(9.0)$ \\
	 & $1.87$ & $4.1 \times 10^{-3}$ & $782.37(28)(6)$ & $2.02(5)(2)$ & $10.1(2.4)(1.4)$ & $494.9(5.3)(3.1)$ \\
	\bottomrule
\end{tabular}}
\caption{Comparison of fits to single experiments with and without a phase $\delta_\eps$ in $\eps_{\omega}$. Note that BESIII has only a few data points in the interference region and hence is not able to put a strong constraint on $\delta_\eps$ (the corresponding line is indicated in gray). The first error is the fit uncertainty, inflated by $\sqrt{\chi^2/\mathrm{dof}}$, the second error is the combination of all systematic uncertainties.}
\label{tab:FitsSingleExperiments}
\end{table}

\begin{table}[t]
\small
\centering
\renewcommand{\arraystretch}{0.95}
\scalebox{0.8}{
\begin{tabular}{l c l l l l l l l l}
\toprule
	 & $\chi^2/$dof & $p$-value & $\mw$ [MeV] & $10^3\times \Re\eps_{\omega}$ & $\delta_\eps$ [${}^\circ$] & $10^{10} \times a_\mu^{\pi\pi}|_{\leq 1\GeV}$\\
	\midrule
	Energy scan w/o SND20 & $1.28$ & $2.1\%$ & $781.75(22)(1)$ & $1.97(4)(2)$ & & $498.5(3.4)(2.6)$\\
	 & $1.05$ & $33\%$ & $782.39(23)(2)$ & $1.93(4)(3)$ & $9.9(1.8)(0.4)$ & $497.3(3.1)(3.9)$\\[0.1cm]
	Energy scan & $1.65$ & $6.3\times10^{-7}$ & $781.74(17)(2)$ & $2.01(3)(3)$ & & $497.4(3.0)(4.4)$\\
	 & $1.19$ & $5.2\%$ & $782.37(16)(3)$ & $1.97(3)(3)$ & $10.1(1.3)(0.7)$ & $496.0(2.6)(5.5)$\\[0.1cm]
	All $e^+e^-$ w/o SND20 & $1.25$ & $1.8\times10^{-5}$ & $781.70(9)(4)$ & $2.02(2)(3)$ & & $494.5(1.5)(2.3)$\\
	 & $1.20$ & $3.3\times10^{-4}$ & $782.10(12)(4)$ & $1.96(2)(2)$ & $4.5(9)(8)$ & $494.2(1.4)(2.1)$\\[0.1cm]
	NA7 + all $e^+e^-$ w/o SND20 & $1.23$ & $3.0\times10^{-5}$ & $781.69(9)(3)$ & $2.02(2)(3)$ & & $494.8(1.4)(2.1)$\\
	 & $1.19$ & $4.8\times10^{-4}$ & $782.09(12)(4)$ & $1.97(2)(2)$ & $4.5(9)(8)$ & $494.6(1.5)(1.7)$\\[0.1cm]
	All $e^+e^-$ & $1.36$ & $1.0\times10^{-9}$ & $781.71(8)(3)$ & $2.02(2)(3)$ & & $495.0(1.4)(2.4)$\\
	 & $1.30$ & $2.3\times10^{-7}$ & $782.09(10)(4)$ & $1.97(2)(2)$ & $4.5(8)(8)$ & $494.6(1.4)(2.1)$\\[0.1cm]
	NA7 + all $e^+e^-$ & $1.34$ & $2.5\times10^{-9}$ & $781.71(8)(3)$ & $2.02(2)(3)$ & & $495.2(1.4)(2.2)$\\
	 & $1.28$ & $4.5\times10^{-7}$ & $782.09(10)(4)$ & $1.97(2)(2)$ & $4.5(8)(8)$ & $494.9(1.4)(1.8)$\\
	\bottomrule
\end{tabular}}%
\caption{Comparison of fits to combinations of experiments with and without a phase $\delta_\eps$ in $\eps_{\omega}$. The first error is the fit uncertainty, inflated by $\sqrt{\chi^2/\mathrm{dof}}$, the second error is the combination of all systematic uncertainties.}
\label{tab:FitsCombinations}
\end{table}

\begin{table}[t]
\small
\centering
\renewcommand{\arraystretch}{0.95}
\scalebox{0.8}{
\begin{tabular}{l c l l l l l l l l}
\toprule
	 & $\delta_\eps$ [${}^\circ$] & SD window & int window & LD window & $10^{10} \times a_\mu^{\pi\pi}|_{\leq 1\GeV}$ \\
	\midrule
	SND06					& 				& $13.9(2)(1)$		& $140.0(2.0)(1.0)$	& $345.8(4.7)(3.0)$	& $499.7(6.9)(4.1)$ \\
							& $8.5(2.3)(0.3)$	& $13.9(2)(1)$		& $139.6(1.8)(1.2)$	& $344.3(4.1)(3.6)$	& $497.8(6.1)(4.9)$ \\[0.1cm]
	CMD-2					&				& $13.9(1)(0)$		& $139.5(1.1)(0.4)$	& $343.6(2.7)(1.8)$	& $496.9(4.0)(2.3)$ \\
							& $11.4(3.1)(1.0)$	& $13.9(1)(1)$		& $139.4(1.0)(0.9)$	& $342.6(2.5)(3.2)$	& $495.8(3.7)(4.2)$ \\[0.1cm]
	BaBar					&				& $14.0(1)(0)$		& $140.6(1.0)(0.5)$	& $347.3(2.2)(1.5)$	& $501.9(3.3)(2.0)$ \\
							& $1.3(1.9)(0.7)$	& $14.0(1)(0)$		& $140.6(1.0)(0.5)$	& $347.3(2.3)(1.3)$	& $501.9(3.3)(1.8)$ \\[0.1cm]
	KLOE$''$					&				& $13.6(1)(1)$		& $137.3(6)(6)$		& $340.9(1.4)(1.2)$	& $491.8(2.1)(1.8)$ \\
							& $6.1(1.7)(0.6)$	& $13.6(1)(0)$		& $137.1(6)(4)$		& $340.2(1.4)(1.3)$	& $490.8(2.0)(1.7)$ \\[0.1cm]
	BESIII					&				& $13.7(1)(0)$		& $138.0(1.4)(0.5)$	& $339.0(3.3)(3.4)$	& $490.8(4.8)(3.9)$ \\
							& \color{gray} $17.6(6.9)(1.2)$ & \color{gray} $13.7(1)(0)$	& \color{gray} $137.8(1.3)(0.4)$		& \color{gray} $338.8(3.1)(2.6)$		& \color{gray} $490.3(4.5)(3.1)$ \\[0.1cm]
	SND20					&				& $13.9(2)(1)$		& $139.4(1.9)(1.5)$	& $340.9(4.6)(7.4)$	& $494.2(6.7)(9.0)$ \\
							& $10.1(2.4)(1.4)$	& $13.8(2)(0)$		& $139.2(1.5)(0.5)$	& $341.9(3.7)(2.6)$	& $494.9(5.3)(3.1)$ \\
	\midrule
	NA7 + all $e^+e^-$ w/o SND20	&				& $13.7(0)(0)$		& $138.3(4)(5)$		& $342.7(1.0)(1.6)$	& $494.8(1.4)(2.1)$ \\
							& $4.5(9)(8)$		& $13.7(0)(0)$		& $138.3(4)(4)$		& $342.5(1.0)(1.3)$	& $494.6(1.5)(1.7)$ \\
	\bottomrule
\end{tabular}}%
\caption{Decomposition of $10^{10} \times a_\mu^{\pi\pi}|_{\leq 1\GeV}$ into the Euclidean windows from Ref.~\cite{Blum:2018mom}. The first error is the fit uncertainty, inflated by $\sqrt{\chi^2/\mathrm{dof}}$, the second error is the combination of all systematic uncertainties.}
\label{tab:FitsEuclideanWindows}
\end{table}

To gauge the impact of a possible phase in $\epsrw$ on the HVP contribution to $a_\mu$, we generalize the global fits from Ref.~\cite{Colangelo:2018mtw}, including a free imaginary part via the prescription~\eqref{eq:GomegaMod}, and express our results in terms of $\Re\epsrw$ and $\delta_\eps$. In particular, we now include the BESIII data~\cite{Ablikim:2015orh} and the SND measurement~\cite{SND:2020nwa}, which became available after Ref.~\cite{Colangelo:2018mtw}.\footnote{In the case of Ref.~\cite{Ablikim:2015orh} the corrected covariance matrix was critical for the inclusion of this data set in a statistically meaningful way. For the uncertainty of the energy calibration at the $\rho$ peak we use $\Delta E=0.6\MeV$~\cite{Redmer} and $\Delta E=0.26\MeV$~\cite{Kupich}, respectively.} The results for the fits are shown in Table~\ref{tab:FitsSingleExperiments} (single experiments) and Table~\ref{tab:FitsCombinations} (combinations), in terms of the most relevant parameters: goodness of fit, the $\omega$ mass, real part and phase of $\epsrw$, and the contribution to $a_\mu$. In Table~\ref{tab:FitsEuclideanWindows}, we also provide the decomposition into the Euclidean windows from Ref.~\cite{Blum:2018mom}.

In most cases, we observe a moderate improvement when a non-vanishing phase is admitted, the main exception being the SND20 data, which we cannot describe with our dispersive representation otherwise. Accordingly, in this case the resulting phase comes out around $10^\circ$ and thus much larger than can be justified via radiative intermediate states. A similarly large phase is also found for the previous energy-scan experiments SND06 and CMD-2, but in these cases good fits can still be found when imposing a realistic size of $\delta_\eps$. Even if a large phase is admitted in the fit to the SND20 data, the fit quality remains rather poor.\footnote{The fit presented in Ref.~\cite{SND:2020nwa} in terms of a sum of Breit--Wigner functions for $V=\rho,\omega,\rho'$ displays a slightly better fit quality, $\chi^2/\text{dof}=47/30=1.57$, with $p$-value of $2.5\%$, but such a representation cannot be reconciled with the analytic properties of the pion VFF.} 
As long as the reason for this behavior, which might point towards underestimated systematic effects, is not understood, we will therefore take the global fit to all experiments apart from SND20 as our new central result, i.e.
\begin{align}
\label{central_results}
 \Re\epsrw&=1.97(3)\times 10^{-3}, & \delta_\eps&=4.5(1.2)^\circ,\notag\\
 a_\mu^{\pi\pi}|_{\leq 1\GeV}&=494.6(2.3)\times 10^{-10}, & \mw&=782.09(12)\MeV.
\end{align}

For BESIII, the preferred central value for $\delta_\eps$ comes out even larger, yet with a very large uncertainty that reflects the limited sensitivity to $\delta_\eps$, resulting from a relatively small number of data points in the $\rho$--$\omega$ region (accordingly, this line is indicated in light gray in Tables~\ref{tab:FitsSingleExperiments} and~\ref{tab:FitsEuclideanWindows}). Finally, the KLOE fits produce a phase slightly larger than expected, while the BaBar data are even consistent with $\delta_\eps=0$. We thus observe a large spread in the results for the phase of $\delta_\eps$, pointing towards systematic differences among the data sets in the $\rho$--$\omega$ region.    

In addition, we confirm a correlation between $\delta_\eps$ and $\mw$, as already observed in Ref.~\cite{Lees:2012cj}: the larger the phase, the larger the extracted value of $\mw$. However, as discussed in more detail in Sec.~\ref{sec:Momega}, the size of the phase permitted by radiative intermediate states, roughly in line with the result of the global fit shown in Table~\ref{tab:FitsCombinations}, does not suffice to remove the tension with $\omega$-mass determinations from $e^+e^-\to 3\pi$ and $e^+e^-\to\pi^0\gamma$.

In this regard, we also observe that the BESIII data suggest larger values of $\mw$ than all other data sets, with the result for $\delta_\eps=0$ close to the global fit with non-vanishing phase. Within uncertainties there is still consistency, but it is noteworthy that the size and direction of the effect echo a similar tension in $e^+e^-\to 3\pi$~\cite{BESIII:2019gjz,BABAR:2021cde,Achasov:2003ir,Akhmetshin:2003zn} and $\eta'\to\pi^+\pi^-\gamma$~\cite{Holz:2022hwz,BESIII:2017kyd}. 

\section{Isospin-breaking contribution to $\boldsymbol{a_\mu}$ from $\boldsymbol{\rho}$--$\boldsymbol{\omega}$ mixing}
\label{sec:IB}

\begin{table}[t]
\small 
\centering
\begin{tabular}{l l l l l}
\toprule
		 & SD window & int window & LD window & total \\
	\midrule
	$10^{10} \times \amurw$, $\delta_\eps=0$ & $0.08(0)(0)$ & $1.06(1)(2)$ & $3.23(3)(5)$ & $4.37(4)(7)$ \\
	$10^{10} \times a_\mu^{\pi\pi,\text{FSR}}$, $\delta_\eps=0$ & $0.11(0)(0)$ & $1.12(0)(0)$ & $3.00(1)(1)$ & $4.23(1)(2)$ \\
	\midrule
	$10^{10} \times \amurw$, $\delta_\eps=4.5(1.2)^\circ$ & $0.05(0)(0)$ & $0.83(5)(4)$ & $2.79(9)(6)$ & $3.68(14)(10)$ \\
	$10^{10} \times a_\mu^{\pi\pi,\text{FSR}}$, $\delta_\eps=4.5(1.2)^\circ$ & $0.11(0)(0)$ & $1.12(0)(0)$ & $3.00(1)(1)$ & $4.24(1)(2)$ \\
	\bottomrule
\end{tabular}
\caption{IB contribution to $a_\mu^{\pi\pi}|_{\leq 1\GeV}$ due to $\rho$--$\omega$ mixing, compared to the effect of FSR and split into the different Euclidean windows from Ref.~\cite{Blum:2018mom}. We only include the linear effects, i.e., $\Order(\epsrw)$ for the $\rho$--$\omega$-mixing contribution and the $\Order(e^2)$ effect for FSR---$\Order(e^2\epsrw)$ effects give very small corrections. The results correspond to the combined fit to all experiments apart from SND20. The first error is the fit uncertainty, inflated by $\sqrt{\chi^2/\mathrm{dof}}$, the second error is the combination of all systematic uncertainties.}
\label{tab:IBcontribution}
\end{table}

Based on the dispersive representation~\eqref{eq:PionVFF} we can quantify $ \amurw$---the IB contribution to $a_\mu$ due to $\epsrw$--- by contrasting the full result to the HVP integral evaluated with $\epsrw=0$. In principle, there is some ambiguity due to final-state radiation (FSR), but in practice this effect comes out well below $0.1\times 10^{-10}$. For definiteness, in Table~\ref{tab:IBcontribution} we show the variant without FSR, to isolate the pure $\Order(\epsrw)$ terms. 

In general, $\amurw$ is sensitive to the assumed line shape~\cite{Wolfe:2009ts}. However, we find that the dispersive representations~\eqref{Gomega} or~\eqref{eq:GomegaMod} are quite robust in that regard, i.e., with the threshold behavior and the properties close to the $\omega$ pole determined, the remaining interpolation only has a marginal effect, e.g., changing $s\to \mw^2$ in the numerator of Eq.~\eqref{gomega} changes the outcome for $\amurw$ again by less than $0.1\times 10^{-10}$. In contrast, whether or not a phase in $\epsrw$ is permitted does change the resulting value for $\amurw$ in a significant way, and we show results for both scenarios (the difference in the FSR contribution is of $\Order(e^2\epsrw)$ and negligible). Since $\epsrw$ in the global fit comes out close to the narrow-resonance expectation, we quote the variant with non-vanishing $\delta_\eps$ as our preferred result, which has already been used as input in estimating the three-flavor quark-disconnected contribution to $a_\mu$ in Ref.~\cite{Boito:2022rkw}.\footnote{The tiny difference to the number for $\amurw$ quoted therein as private communication originates from the improved implementation of the $\pi^0\gamma$ threshold~\eqref{eq:GomegaMod}.} 
Finally, we also provide the breakdown of $\amurw$ onto the Euclidean windows from Ref.~\cite{Blum:2018mom}. 

Further, for the comparison to lattice QCD it is also of interest to study
the decomposition of $\epsrw$ into its $\Order(e^2)$ and $\Order(m_u-m_d)$ pieces, as was discussed in the context of resonance chiral perturbation theory in Ref.~\cite{Urech:1995ry}. Translated to our normalization one has the prediction~\cite{Urech:1995ry}
\beq
\label{eps_RChT}
\tilde\eps_\omega = \frac{2}{3R}\frac{M_{K^*}-M_V}{M_V}-\frac{e^2}{|g_{\omega\gamma}|^2},\qquad R=\frac{m_s-\hat m}{m_d-m_u},\qquad \hat m = \frac{m_u+m_d}{2},
\eeq
where we have written the electromagnetic component in terms of the $\omega$--$\gamma$ coupling, as this is the quantity that enters directly in the corresponding diagram. However, this latter diagram produces a one-particle-reducible correction, and would thus be subtracted when vacuum polarization is removed from the $e^+e^-\to\pi^+\pi^-$ cross sections. Accordingly, we have that our conventions are related to the ones of Ref.~\cite{Urech:1995ry} 
by
\beq
\tilde \eps_{\omega}=\epsrw+\tilde\eps_\omega|_{e^2},\qquad  \tilde\eps_\omega|_{e^2}=-\frac{e^2}{|g_{\omega\gamma}|^2}=-0.34(1)\times 10^{-3},
\eeq
using $\Gamma[\omega\to e^+e^-]={4\pi\alpha^2M_\omega}/{(3|g_{\omega\gamma}|^2)}$ and the average of Ref.~\cite{ParticleDataGroup:2020ssz} for the $\omega\to e^+e^-$ branching fraction.

The prediction for the $\Order(m_u-m_d)$ part of $\tilde \eps_{\omega}$, which coincides with our $\epsrw$, is far less robust, as already from higher-order quark-mass and $SU(3)$-breaking corrections one would expect an accuracy around $30\%$. Using the $N_f=2+1$ and $N_f=2+1+1$ averages from Ref.~\cite{Aoki:2021kgd}, $R=38.1(1.5)$~\cite{RBC:2014ntl,Durr:2010vn,Durr:2010aw,MILC:2009ltw,Bruno:2019vup,Fodor:2016bgu} and $R=35.9(1.7)$~\cite{EuropeanTwistedMass:2014osg,Bazavov:2017lyh,FermilabLattice:2014tsy,Giusti:2017dmp,MILC:2018ddw}, respectively, and identifying the vector mesons with the neutral $\rho$ and $K^*$ resonances, the predictions for the strong IB contribution become $\tilde\eps_\omega|_{m_u-m_d}^{2+1}= 2.71(11)\times 10^{-3}$, $\tilde\eps_\omega|_{m_u-m_d}^{2+1+1}= 2.88(14)\times 10^{-3}$, about $40\%$ larger than results from our fit to the $e^+e^-\to\pi^+\pi^-$ data.
This is in line with subsequent work on vector mesons in chiral perturbation theory~\cite{Bijnens:1996nq,Bijnens:1997ni}, which concluded that higher-order corrections can be substantial. This includes photon loops, short-distance corrections, and meson loops, parts of which scale with $e^2$ and thus lead to electromagnetic effects not subtracted when removing vacuum polarization in the definition of the bare cross section.   

However, from the LO expression~\eqref{eps_RChT} it still follows that the $\rho$--$\omega$-mixing contribution to $a_\mu$ should be considered primarily a quark-mass effect,  
\beq
\amurw\big[e^2, \text{LO}\big]=0,\qquad 
\amurw\big[m_u-m_d, \text{LO}\big]=3.68(17)\times 10^{-10},
\eeq
which is expected to yield the dominant strong IB contribution to $a_\mu$.\footnote{Resonance-enhanced threshold effects in the $\bar K K$ channels largely cancel between $K^+K^-$ and $K_SK_L$.} This number agrees well with a recent estimate from $SU(3)$ chiral perturbation theory, $a_\mu[m_u-m_d]\big|_\text{\cite{James:2021sor}}=3.32(89)\times 10^{-10}$, where the required low-energy constant is determined from hadronic $\tau$ decays. Both indicate a somewhat larger central value than the lattice-QCD result $a_\mu[m_u-m_d]\big|_\text{\cite{Borsanyi:2020mff}}=1.9(1.2)\times 10^{-10}$.

\section{Consequences for the $\boldsymbol{\omega}$ mass}
\label{sec:Momega}

\begin{table}[t]
\small 
\centering
\begin{tabular}{l l l l l}
\toprule
	Reference & $e^+e^-\to 3\pi$ & $e^+e^-\to\pi^0\gamma$ & $e^+e^-\to 2\pi$ & PDG average\\
	\midrule
Ref.~\cite{Hoferichter:2019gzf} & $782.631(28)$ & & &\\
Ref.~\cite{Hoid:2020xjs} &  & $782.584(28) $ & &\\
Ref.~\cite{Colangelo:2018mtw} &  & & $781.68(10)$ &\\
This work, $\delta_\eps=0$ &  & & $781.69(9)$ &\\
This work, $\delta_\eps=4.5(1.2)^\circ$ &  & & $782.09(12)$ &\\\midrule	Ref.~\cite{ParticleDataGroup:2020ssz} & & & & $782.53(13)$\\
	\bottomrule
\end{tabular}
\caption{Dispersive determinations of $\mw$ in MeV from $e^+e^-$ reactions, compared to the global average from Ref.~\cite{ParticleDataGroup:2020ssz}. In all cases, vacuum-polarization corrections are not included, and the average from Ref.~\cite{ParticleDataGroup:2020ssz} has been adjusted accordingly using $\Delta \mw=0.13\MeV$~\cite{Holz:2022hwz}.}
\label{tab:Momega}
\end{table}

The correlation between $\delta_\eps$ and $\mw$ discussed in Sec.~\ref{sec:fits} affects the resulting determination of $\mw$ from $e^+e^-\to 2\pi$. In Table~\ref{tab:Momega} we compare our updated extraction from the $2\pi$ data to analogous ones from $e^+e^-\to 3\pi$, $e^+e^-\to\pi^0\gamma$, as well as the average from Ref.~\cite{ParticleDataGroup:2020ssz}.  As discussed in more detail in Refs.~\cite{Colangelo:2018mtw,Hoferichter:2019gzf,Hoid:2020xjs}, the PDG average involves a cancellation between determinations from $e^+e^-\to\pi^0 \gamma$~\cite{CMD-2:2004ahv} and $\bar p p\to \omega\pi^0\pi^0$~\cite{CrystalBarrel:1993gtk}, while dominated by Breit--Wigner-based extractions from $e^+e^-\to 3\pi$~\cite{Achasov:2003ir,Akhmetshin:2003zn} that are in agreement with the dispersive result given in Table~\ref{tab:Momega} (further confirmed by the recent BaBar measurement~\cite{BABAR:2021cde}, while BESIII suggests a larger value~\cite{BESIII:2019gjz}). Our updated value for $\delta_\eps=0$ changes only marginally compared to Ref.~\cite{Colangelo:2018mtw}, leading to the same $5\sigma$ tension with the PDG value observed therein. Allowing a finite value for $\delta_\eps$ instead removes about half the discrepancy, but we emphasize that this effect cannot explain the entire tension as it would require a size of $\delta_\eps$ that cannot be reconciled with the strength of the radiative channels giving rise to a phase in $\epsrw$ in the first place.

\section{Conclusions}
\label{sec:conclusions}

In this work we performed a detailed study of $\rho$--$\omega$ mixing in $e^+e^-\to\pi^+\pi^-$, based on a dispersive representation of the pion vector form factor. In particular, we investigated the role of imaginary parts that can be generated by radiative intermediate states coupling $\omega$ and $\rho$ resonances, estimated their size by narrow-width arguments, and devised a strategy to include their effect in fits to the $e^+e^-\to\pi^+\pi^-$ data base. We found that while the size of the phase of the $\rho$--$\omega$ mixing parameter in a global fit does come out in agreement with narrow-resonance expectations, see Eq.~\eqref{central_results} for the central results, there is a substantial spread among the different data sets, ranging from a vanishing phase to values as large as $10^\circ$. As applications, we derived the isospin-breaking part of the HVP contribution to  $a_\mu$ originating from $\rho$--$\omega$ mixing and quantified the changes in the extracted value of the $\omega$ mass when a non-vanishing phase is permitted. 

Our work reveals systematic differences in the low-energy hadronic cross sections that go beyond the well-known BaBar--KLOE tension in the $e^+e^-\to\pi^+\pi^-$ total cross section, including the spread in the phase of the $\rho$--$\omega$ mixing parameter and discrepancies in the $\omega$ mass extracted from different decay channels, both of which can be unambiguously defined in terms of pole parameters and residues. While of course resolving the tension in the HVP integral itself carries the highest priority, forthcoming high-precision data on $e^+e^-\to\pi^+\pi^-$ should also allow one to address the tensions pointed out here, and thus increase confidence that the hadronic cross sections are understood at the level required for robust data-driven evaluations of the HVP contribution to the anomalous magnetic moment of the muon.

\acknowledgments  
We thank Christoph Hanhart for valuable discussions that led to the derivation in Sec.~\ref{sec:radiative}, and Kim Maltman for encouragement to write up these results. 
We further thank Pablo S\'anchez-Puertas for pointing out the issue of one-photon-reducible contributions to $\epsrw$, and Hans Bijnens for discussions on $\epsrw$ in resonance chiral perturbation theory. 
Financial support by the DFG through the funds provided to the Sino--German Collaborative
Research Center TRR110 ``Symmetries and the Emergence of Structure in QCD''
(DFG Project-ID 196253076 -- TRR 110) and the SNSF (Project Nos.\ 200020\_175791, PCEFP2\_181117, and PCEFP2\_194272) is gratefully acknowledged. 


\bibliographystyle{apsrev4-1_mod}
\bibliography{ref}
	
\end{document}